\documentclass[conference]{IEEEtran}
\IEEEoverridecommandlockouts
\usepackage{cite}
\usepackage{amsmath,amssymb,amsfonts}
\usepackage{algorithmic}
\usepackage{graphicx}
\usepackage{textcomp}
\usepackage{xcolor}
\usepackage{hyperref}
\usepackage{caption}
\usepackage{subcaption}
\usepackage{float}
\hypersetup{colorlinks,linkcolor={blue},citecolor={blue},urlcolor={blue}}
\def\BibTeX{{\rm B\kern-.05em{\sc i\kern-.025em b}\kern-.08em
    T\kern-.1667em\lower.7ex\hbox{E}\kern-.125emX}}
\usepackage{algorithm2e}
\begin{document}

\title{Exploring DRAM Cache Prefetching for Pooled Memory}

\author{
\IEEEauthorblockN{1\textsuperscript{st} Chandrahas Tirumalasetty} 
\IEEEauthorblockA{\textit{Dept. of Electrical \& Computer Engineering} \\
\textit{Texas A\&M University}\\
College Station, TX USA \\
chandrahas996@tamu.edu}
\and
\IEEEauthorblockN{2\textsuperscript{nd} Narasimha Reddy Annapareddy}
\IEEEauthorblockA{\textit{Dept. of Electrical \& Computer Engineering} \\
\textit{Texas A\&M University}\\
College Station, TX, USA \\
reddy@tamu.edu}
}
\maketitle
\begin{abstract}
Hardware based memory pooling enabled by interconnect standards like CXL have been gaining popularity amongst cloud providers and system integrators.  While pooling memory resources has cost benefits, it comes at a penalty of increased memory access latency. With yet another addition to the memory  hierarchy, local DRAM can be potentially used as a block cache(DRAM Cache) for fabric attached memory(FAM) and data prefetching techniques can be used to hide the FAM access latency. This paper proposes a system for prefetching sub-page blocks from FAM into DRAM cache for improving the data access latency and application performance. We further optimize our DRAM cache prefetch mechanism through enhancements that mitigate the performance degradation due to bandwidth contention at FAM. We consider the potential for providing additional functionality at the CXL-memory node through weighted fair queuing of demand and prefetch requests. We compare such a memory-node level approach to adapting prefetch rate at the compute-node based on observed latencies.  We evaluate the proposed system in single node and multi-node configurations with applications from SPEC, PARSEC, Splash and GAP benchmark suites. Our evaluation suggests DRAM cache prefetching result in 7\% IPC improvement and both of proposed optimizations can further increment IPC by 7-10\%. 
\end{abstract}
\section{Introduction}
Modern workloads are evolving rapidly.  Heavy use of ML techniques, from the data-center to client/mobile, are placing new, more stringent demands on system design across all platforms.  Many of these ML
techniques from large language models
\cite{attention-is-all-you-need,bert,Colbert,llama} to video processing 
\cite{video-edge,drizzle,Chameleon} and others rely on
large amounts of data for training and sometimes for retrieval. In the data center, where a given workload may reference Terabytes of data, spread across many nodes, great demands on computer memory systems are
becoming common place~\cite{ferdman2012clearing,dsservers}.  
%The trend towards server virtualization \cite{barham2003xen,wang2010impact}, is
%furthering this pressure on computer memory systems.  %These demands continue to accelerate faster than the resources in computer systems and as a result are forcing algorithm designers and system designers to devise efficient data movement and consumption as a top priority.

In recent years, the performance gap between DRAM and disk has grown so large to lead system designers to eschew using storage to extend DRAM entirely in favor of over-provisioning
DRAM~\cite{inmemDAC2020,inmemglsvlsi2020,memcached,resilientDB,sparkbench,ramcloud}. Many different techniques have been proposed to reduce the cost of
data movement and page fault handling penalties. These range from employing a larger DRAM memory and running applications entirely in memory \cite{ramcloud,log-dram-fast14}, prefetching data blocks \cite{span,Jain_Lin,Kim:2016}, 
%3D stacked memories \cite{Loh:2008,sun2009novel,zhang2009exploring} 
 and employing remote memories \cite{infiniswap,leap,fastswap}.

Datacenter servers host applications with diverse memory requirements. Provisioned larger DRAM capacities can potentially ameliorate the performance needs of some applications, while not being touched by rest. Hence, \textit{Memory Underutilization} is rampant in today's datacenters. Trace analysis from production clusters at Google and Alibaba revealed that 45\%-60\% of allocated memory to jobs is not utilized \cite{Legoos}. Untouched memory for virtual machine instances (VM) in Azure servers, average about 25\%, while full of compute is used \cite{pond}.

Furthermore, \$/GB price of DDR memory has plateaued for last few generations \cite{memory_cost_2}. As a result, cost to provision larger memory capacities in today's severs is steeply increasing with every generation. Memory contributes 37\% -50\% of total cost of ownership(TCO) of server fleet \cite{tpp, azure_memory_cost}. Memory underutilization incurs substantial costs at the scale of today's datacenters.

\textit{Memory Disaggregation} enables applications to avail memory from a central resource on an ad-hoc basis, freeing the memory from being tied up statically at node-level. Modern data center servers have been exploring the potential of disaggregated memories to provide a less expensive means of furnishing memory\cite{far-memory-warehouse, Heteroos,Legoos}. The disaggregated
approaches have taken two parallel paths: (1) RDMA based approaches that employ memory at another node as remote/far memory, accessed through Operating System (OS) based paging mechanisms~\cite{infiniswap, leap, fastswap}.
(2) The second approach plans to employ CXL to provide a shared common pool of memory across multiple nodes~\cite{CXL,Redy-cache,Thermostat,rethinking-disaggregated}. We refer such memory organization as Fabric Attached Memory(FAM). With either approach, it is expected that the memory is used more efficiently across different workloads with divergent memory needs.
%The disaggregation is expected to lead to smoothing of memory usage across many workloads and less stranding of memory resources.  
%Several studies have shown that even a bit of resource sharing can lead to improved efficiencies~\cite{little-flexibility,resource-pooling}.

These architectural paths result in additional layers in the cache-memory hierarchy, with remote or shared common DRAM across an interconnect being the new layer beyond the DRAM.
As new memory layers including disaggregated memories, far memories and non-volatile memories close the gap in speed between different layers of memory, it has become necessary to pursue lower latency approaches for accessing data from these new layers of memory \cite{fastswap,johnnycache}. This paper pursues the approach of prefetching data between DRAM and the lower layers of memory such as disaggregated memory (over CXL-like interconnect) to this end. 

This paper explores the potential of utilizing LLC misses that are visible at the root-complex level to build a prefetching mechanism between FAM and DRAM, utilizing a portion of local DRAM as a hardware managed cache for FAM. We call this proposed cache as DRAM cache. We employ SPP \cite{spp} as an example prefetcher to demonstrate the performance gains with DRAM cache prefetching, but other prefetchers can be employed as well. Our DRAM cache prefetcher maintains the metadata for the cached FAM data. Root complex equipped with a prefetcher redirects requests to cached data to the DRAM cache. On a hit, the cached data will see DRAM latencies instead of FAM latencies.
Unlike previous mechanisms that considered page level transfers between DRAM and lower layer memory 
\cite{fastswap,leap,johnnycache,CXL-2LM}
%\cite{feiwen-hybrid, fastswap,leap}, 
we consider the potential for sub-page level prefetches at the hardware level. 

Since multiple nodes can pool memory from FAM, it is imperative that the FAM bandwidth is utilized and shared across multiple nodes effectively. As previous work has shown, prefetch throttling \cite{NST,fdp} is an effective mechanism to utilize the memory bandwidth well. We take inspiration from this earlier work and incorporate ideas for prefetch throttling to effectively manage the FAM bandwidth across demand and prefetch streams across multiple nodes. We take inspiration from network congestion algorithms \cite{RED} to develop bandwidth adaptation techniques at the source(compute node). Since CXL-connected memory devices can be enhanced with extra functionality, we evaluate the potential of employing Weighted Fair Queueing (WFQ) at the memory node and compare that approach with prefetch throttling at the source. 

This paper makes the following significant contributions:
\begin{itemize}
   \item Proposes a system architecture for caching and prefetching FAM data at local DRAM. Cache being managed at the granularity of sub-page blocks.
    \item Proposes an adaptive prefetching mechanism that throttles DRAM cache prefetches in response to congestion at FAM.
    \item Proposes a WFQ-enabled CXL-memory node and compares its performance against prefetch throttling at the source.
    \item Evaluates the proposed prefetch mechanism to demonstrate its efficacy in single node and multiple node configurations.
\end{itemize}
\section{Background}
\subsection{CXL enabled memory pooling}
Compute Express Link(CXL) \cite{cxl_specification} is a cache-coherent interconnect standard for processors to communicate with devices like accelerators and memory expanders. CXL builts upon the physical layer of PCIe(electrically compatible). CXL offers 3 kinds of protocols- CXL.cache, CXL.mem, CXL.io. Any device that connects to the host processor using CXL can use either or all of aforementioned protocols. CXL identifies 3 types of devices that use one or more these protocols. Type-1 devices like Network Interface Controller(NIC) has a cache hierarchy but does not have local memory, uses CXL.cache. Type-2 device like GPU, FPGA which comprise both caches and local memory uses CXL.cache, CXL.mem. Type-3 device like memory expanders that does not have a local cache hierarchy use CXL.mem. 
%Intel \& AMD have began supporting CXL in their datacenter processor offerings, while memory vendors like Samsung, and Micron are bringing CXL-compatible memory devices to the market. 
%\textcolor{green}{CT: Might need citations regarding support for CXL, or remove the sentence completely}

\begin{figure}[h]
    \centering
    \includegraphics[scale=0.23]{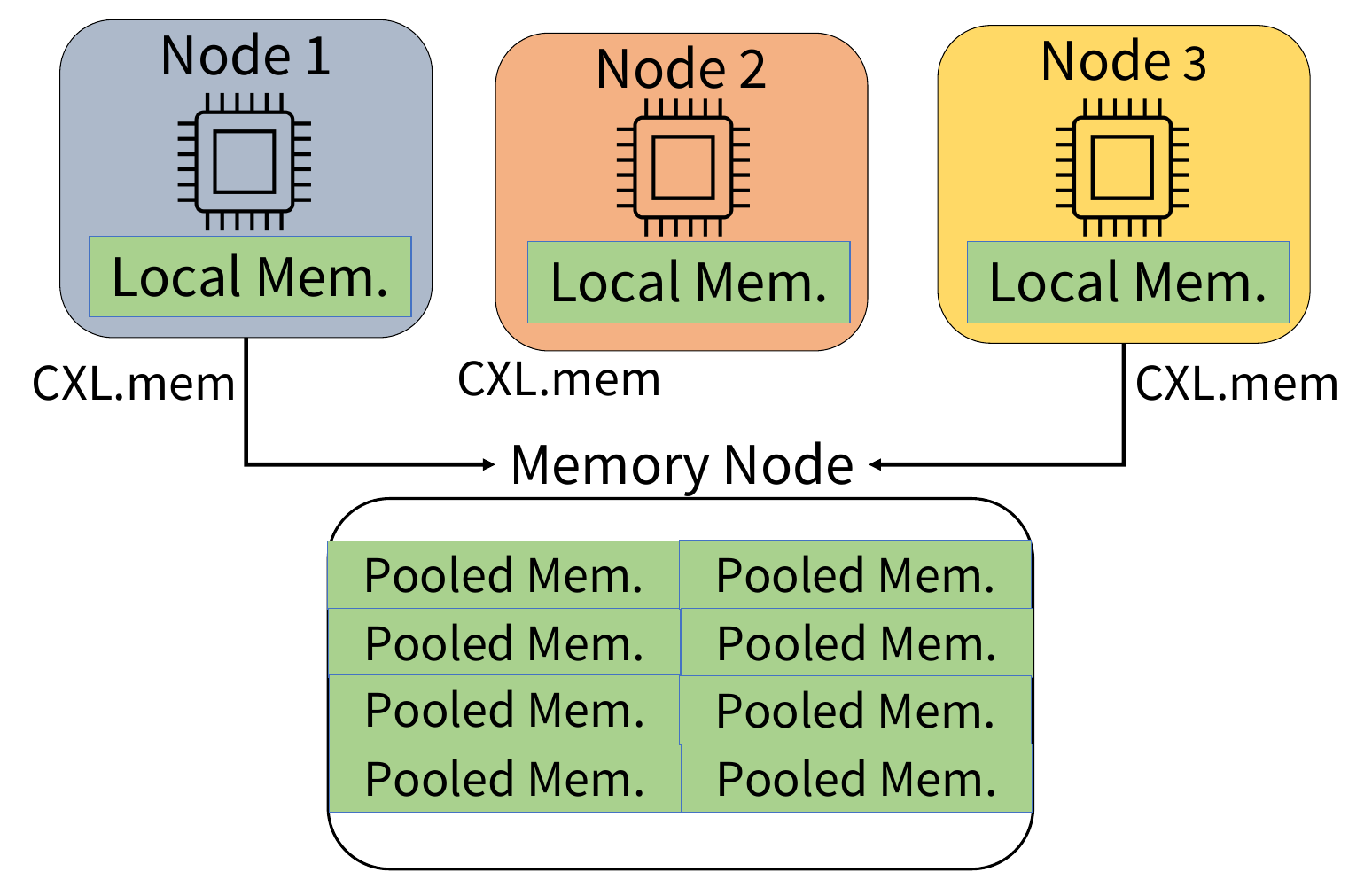}
    \caption{CXL.mem enabled memory pooling}
    \label{fig:multinode)_cxl_mempool}
    \vspace{-1.5em}
\end{figure}

\begin{figure}[h]
    \centering
    \includegraphics[scale=0.23]{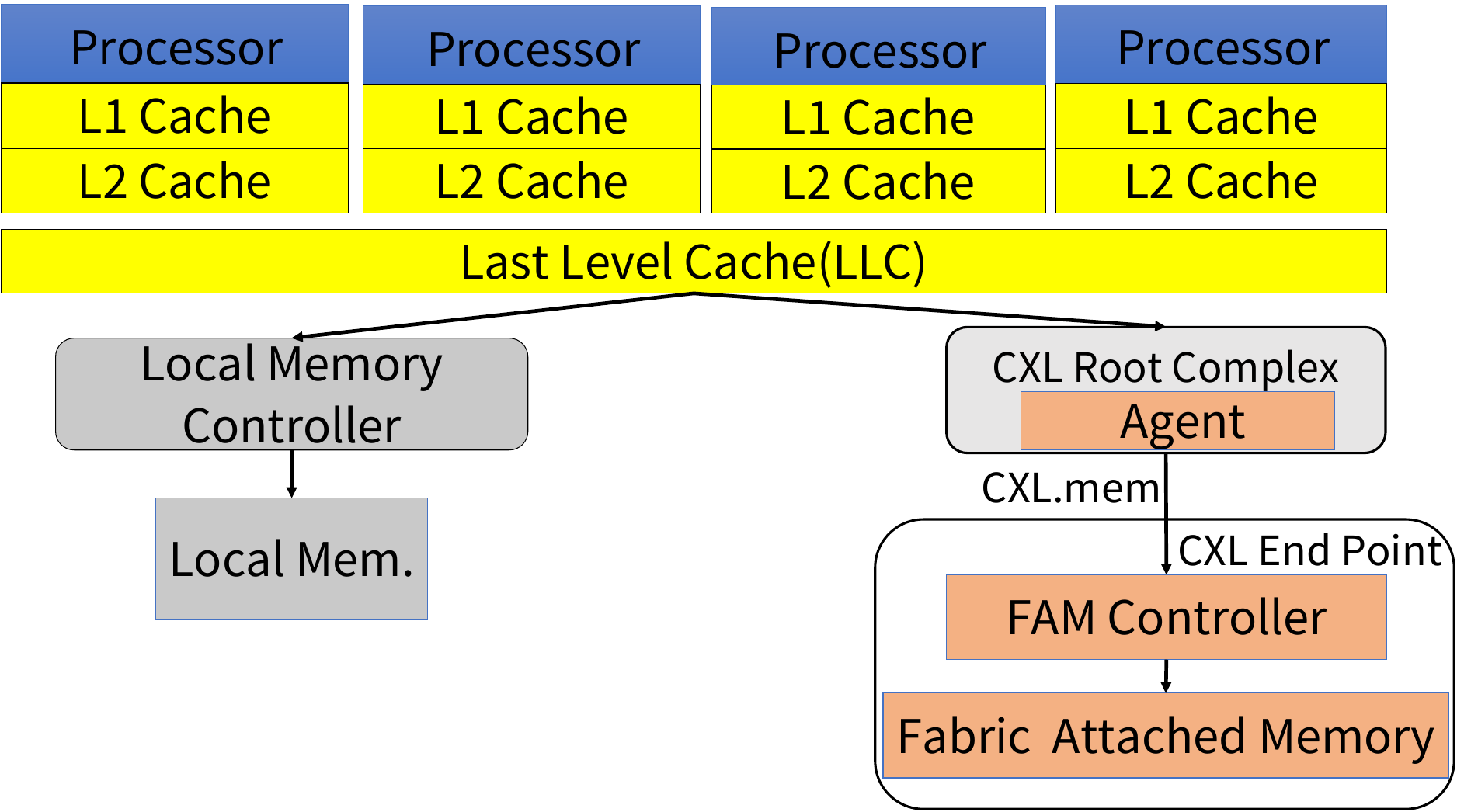}
    \caption{CXL \& Fabric Attached Memory(FAM) Architecture}
    \label{fig:cxl_fam}
\end{figure}

Our discussion in this paper is based on systems that leverages CXL.mem protocol for memory pooling. Fig. \ref{fig:multinode)_cxl_mempool} shows compute nodes pooling memory resources from a shared memory node. We call the memory attached to the processor using CXL as Fabric Attached Memory(FAM).   
Fig.\ref{fig:cxl_fam} details our system architecture with CXL and FAM components. CXL root complex comprises of an agent that implements CXL.mem protocol. Agent acts on behalf of the host(CPU) handling all communication and data transfers with the CXL end point. In our system CXL end point comprises of FAM device and FAM Controller. FAM Controller directly interfaces with the agent, translating CXL.mem commands into requests that can be understood by the FAM device(eg: DDR commands).

As illustrated, load misses and writebacks from LLC are handled either by the local memory controller or CXL root complex based on the physical addresss. The address decoding is implemeted in Host managed Device Memory(HDM) decoders. During the device enumeration phase, HDM decoders are programmed for every CXL.mem device and their contribution to flat address space of the processor.
%In case the CXL root complex needs to handle a request, the agent is delegated to handle the request. 
\vspace{-1.5em}
\subsection{Memory Prefetching \& Signature Path Prefetcher(SPP)}
Data prefetching techniques to memory hide access latency across different levels of memory hierarchy, are well studied in the literature. Prefetchers typically use learning based approach to predict the future memory access addresses\cite{Kim:2016, Bingo-prefetcher}. Most common features include address delta correlation, program counter(PC) that cause cache misses, and access history.  Recent work has applied sophisticated mechanisms like neural networks\cite{neural-prefetching}, reinforcement learning \cite{Pythia,neural-prefetching} to prefetching.

\subsubsection{Signature Path Prefetcher(SPP)}
\begin{figure}[h]
    \centering
    \includegraphics[scale=0.24]{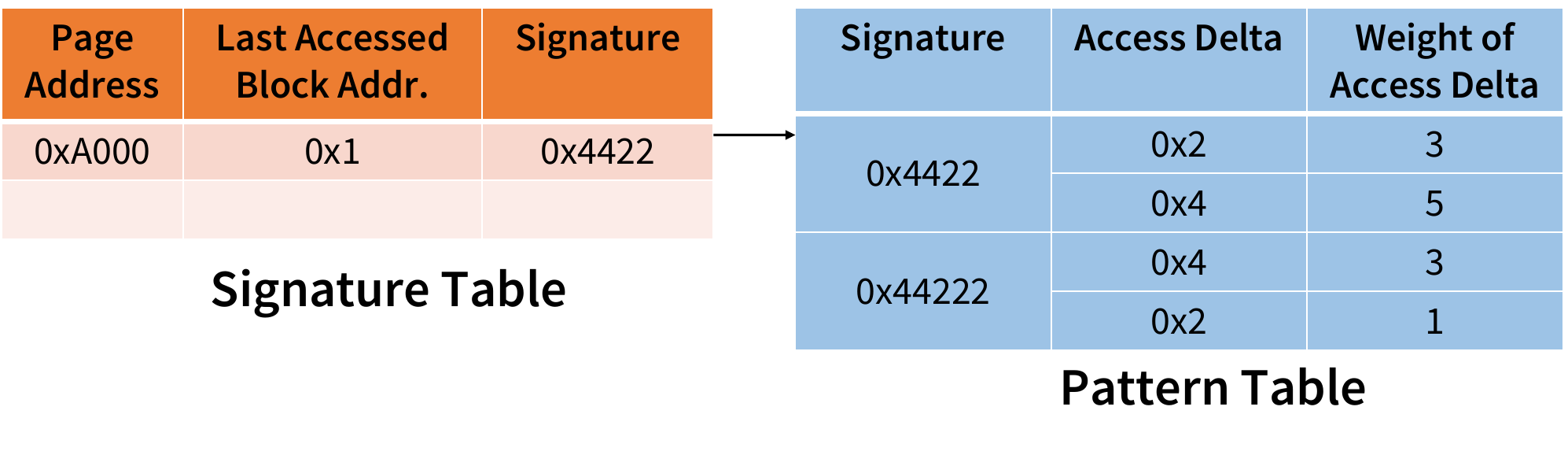}
    \vspace{-1em}
    \caption{Signature and Pattern tables of SPP}
    \label{fig:spp_design}
    \vspace{-1.5em}
\end{figure}

In this work, we use SPP \cite{Kim:2016} as base architecture for our DRAM cache prefetcher. SPP uses signatures to keep track of memory access patterns of the application. Signatures are a compact representation of history of memory access delta's of the program. Architecturally SPP comprises of 2 tables - Signature Table, Pattern Table. Fig. \ref{fig:spp_design} shows the organization of SPP with these tables.
\vspace{-1.5em}

On a cache miss, the page physical address of the cache miss is used to index into the signature table. Output from the signature table gives the last cache miss address(within the same physical page), and the current signature. With this state, we can calculate the delta and updated signature as per formula shown below.
\begin{align*}
 &delta = (Miss\ Address_{current} - Miss\ Address_{previous})\\
 &signature = (signature<<4)\oplus delta
\end{align*}
Now the generated signature is used to index into the pattern table. Pattern table maps signature to address delta's of future memory accesses. Each pattern table entry has the following entries.
\begin{enumerate}
    \item Signature -  Obtained from the signature table. Serves as an index to this table.
    \item Signature weight - Counts the number of times the corresponding signature has been accessed since the creation of entry.
    \item  delta, weight [4] - Address delta that comprise the signature and their corresponding access count. 4 ordered pairs.
\end{enumerate}
The obtained address delta can then be combined with current signature to generate a speculative signature(using the aforementioned formulation). Speculative signature can further be used to index into the signature table, to generate another address delta. This recursive indexing into the pattern table can be continued desired number of times or till the pattern table was not able to provide any more delta's. 

\begin{figure}[h]
    \centering
    \includegraphics[scale=0.26]{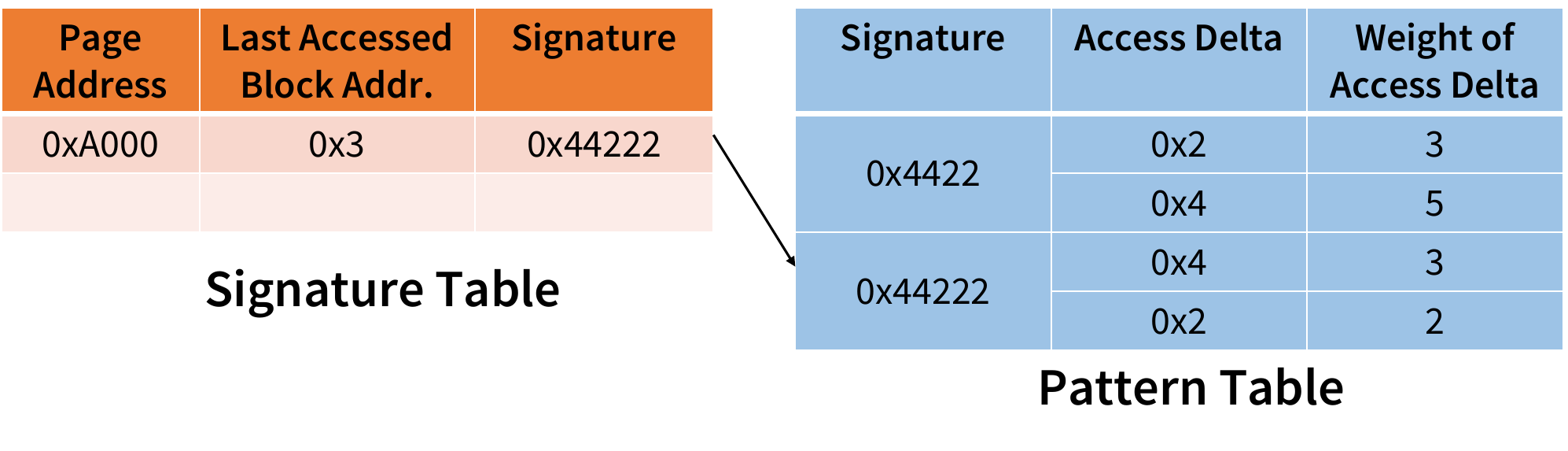}
 
    \caption{State of SPP after access at an address 0xA003}
    \label{fig:spp_design_after_update}
\end{figure}

On an access to the prefetcher(cache miss to a certain page), generated signature and the block address of the current access are used to update the state of SPP. Fig. \ref{fig:spp_design_after_update} shows the state of SPP after an example memory access. Additionally, SPP maintains global history table that bootstraps the learning of access history, when the data access stream moves from one page to another. 
\vspace{-1em}
\vspace{-0.5em}

We note that our proposals in this paper are not specific to SPP prefetcher and ideas from other prefetchers such as \cite{Berti,DSPatch} can be employed with suitable modifications in our system. Our focus in this work is to demonstrate the usefulness of sub-page level prefetching to hide FAM access latency and to present adaptive optimizations to cater to the shared nature of FAM.
\section{System Architecture}
In this section we describe the system architecture components that implement DRAM prefetching/caching mechanism for FAM bound requests. Through the rest of the section, the demands and prefetches that we refer to, are LLC misses and DRAM cache prefetches. DRAM cache prefetches should not be confused with prefetch requests issued by per-core cache prefetcher(core prefetches). Our system architecture doesn't distinguish between type of requests that miss in LLC, hence core prefetch requests that miss in LLC are treated like demand misses and are subsequently used for training the DRAM cache prefetcher as well. 
% Chandrahas: Using red to easily distinguish changes made by me
\subsection{Enhanced Root Complex}
\label{subsec:prefetcher}
\begin{figure}[h]
    \centering
    \includegraphics[scale=0.23]{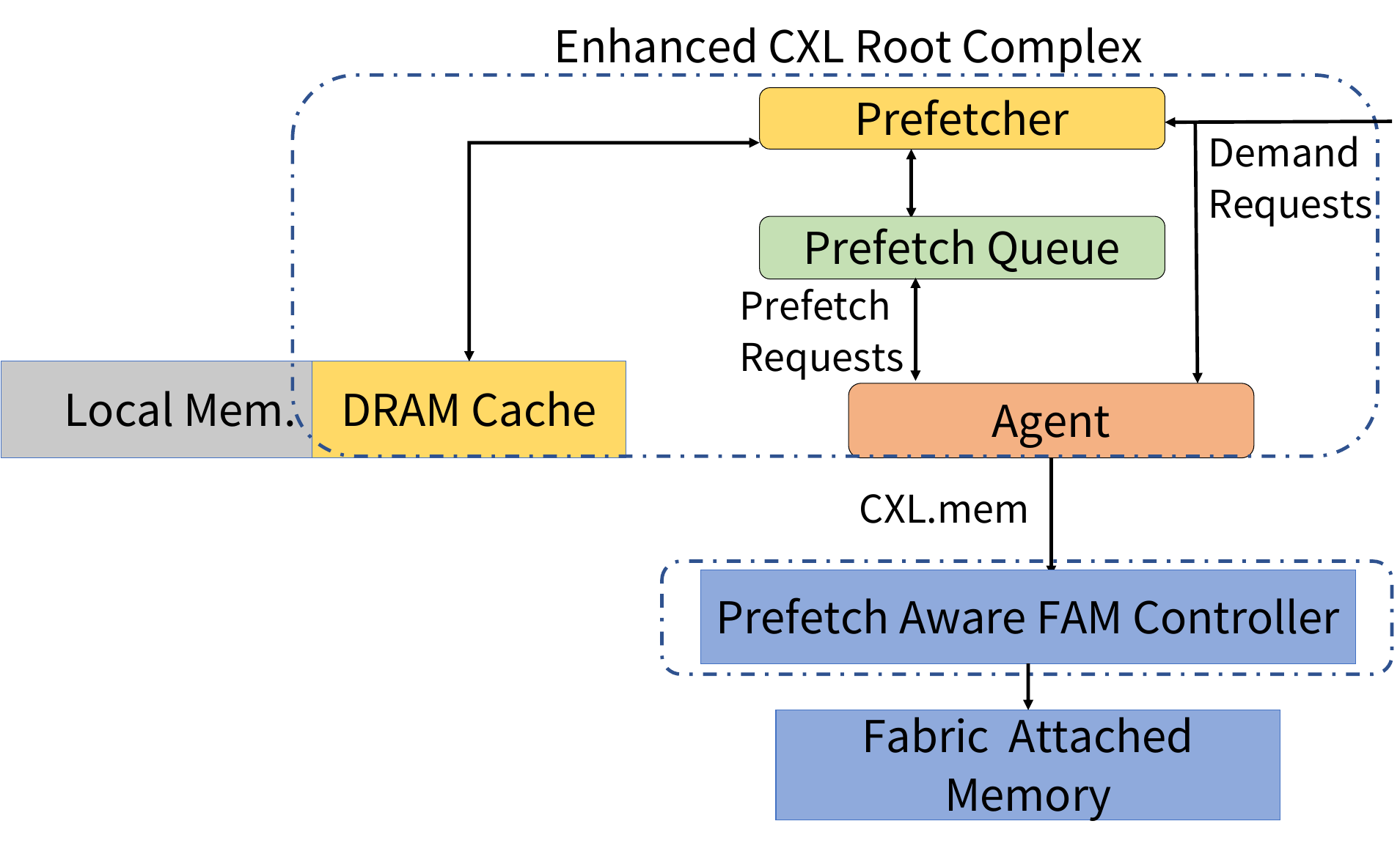}
    \caption{System architecture of root complex with DRAM cache and prefetcher}
    \label{fig:enhanced_sys_arch}
     \vspace{-1em}
\end{figure}

%In this section we discuss the system architecture that serves as baseline for the BLAP optimizations. 
DRAM caching/prefetching is implemented through enhancements the root complex. We add prefetcher and prefetch queue, to facilitate issue of  both prefetch and demand requests to FAM. Fig. \ref{fig:enhanced_sys_arch} outlines the architecture of enhanced root complex. We explain the significance each of the component in detail below.
\vspace{-1em}
\subsubsection{Prefetcher}
As mentioned before, our DRAM cache employs an SPP based prefetcher. We make design changes to SPP to operate with sub-page blocks, instead of 64 byte cache blocks. Our prefetcher trains on node physical address of LLC misses. Based on observed patterns, prefetcher generates addresses that are aligned with sub-page block size. 
%SPP was originally designed to be used in L2 cache of the processor, hence has relatively smaller hardware overhead. Since, we are using it for LLC prefetching, we have more hardware headroom. Hence we doubled the entries in the signature, pattern and global history tables, resulting in double the hardware complexity of original design.
Note that, for memory access to complete, node physical addresses need to be translated to FAM local address. We assume that this translation is handled by the elements lower in the memory hierarchy. Storage overhead required to implement our prefetch is 11kB(2$\times$ to that SPP \cite{spp}). 
\subsubsection{Prefetch Queue}
Along with the prefetcher, we added a fixed-length prefetch queue to the root complex. For every read miss in LLC that is headed to FAM, the prefetcher generates a predefined number of prefetch requests(at the maximum), we call this number the prefetch degree. For each such  prefetch request to be issued to FAM, it should have a vacant position to hold in the prefetch queue. The prefetch request will be held in the queue until the respective response is received. Since, the prefetch queue houses the prefetch requests in progress, the queue provides an easy way to check if a demand request address belongs to any prefetch in progress. In a sense, prefetch queue functionality is similar to MSHR(Miss Status Handling Register) in processor caches. When the prefetch queue is full, no further prefetches are issued until a prefetch response is received. 
%\textcolor{red}{What do we do if the prefetch queue is full? We don't issue that prefetch request? As prefetch requests take longer, does this become a rate limiter?}

We should note that the prefetch queue itself could control the rate of prefetch requests issued to FAM, due to its fixed length. As we will show later, such static approaches work well for a few applications while leading to wasteful prefetching for several applications. Prefetch bandwidth optimizations adapt the prefetch issue rate beyond the fixed length queue.

Prefetchers that fetch data into the on-chip processor caches, share queue with the demand requests in MSHR's. In our design, DRAM cache prefetcher cannot use LLC MSHR due to the difference in block size. While it is possible to use multiple entries in the LLC MSHR, it is not a resource efficient approach.

%We should note that the prefetch queue itself could control the rate of prefetch requests issued to FAM, due to its fixed length. As we will show later, such static approaches work well for a few applications while leading to wasteful prefetching for several applications. BLAP optimizations adapt the prefetch issue rate beyond the fixed length queue.

Prefetch requests that are leaving the prefetch queue are tagged. Architectural components in the fabric or at the FAM node, can likely take advantage of this to enforce priority/QoS schemes. 

\subsection{DRAM Cache}
\label{subsec:dram_cache}
DRAM cache is explicitly managed in hardware without intervention of the operating system(OS). OS, specifically the memory allocator only play a role during initialization phase. The memory allocator should partition the local memory physical address space and expose a contiguous physical address range to be used as a DRAM cache. We assume that such support in the OS already exists.
%One potential straight forward implementation would be to remove certain physical address range from legal physical address range for user space pages and provide it for DRAM cache. 

In this implementation, we manage the DRAM cache as a set-associative cache, with replacement policy being LRU. The meta-data to implement the DRAM cache lookup and replacement will be stored outside the DRAM cache, in the prefetcher state(SRAM buffers). The handling of DRAM cache meta data will be discussed later in this section. 
\begin{figure}[h]
    \centering
    \includegraphics[scale=0.23]{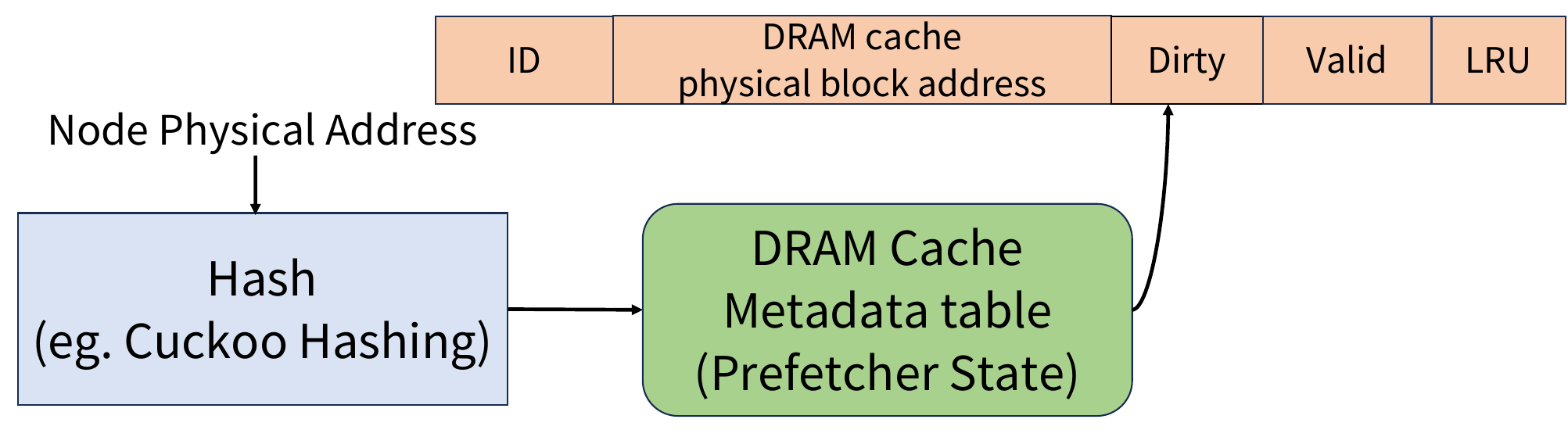}
    \caption{DRAM Cache Metadata format and retrieval}
    \label{fig:dramcache_metadata}
    \vspace{-1em}
\end{figure}
Since FAM address space can be large compared to the DRAM cache, we manage DRAM cache metadata by hashing the FAM addresses into a small number of slots. The number of slots can be higher than the number of available FAM blocks to reduce the probability of collision with another address or it is possible to employ techniques such as Cuckoo hashing \cite{cuckoo}. Like in cpu caches, tag comparison ensures the correctness of hashing, in an event of collision. The allocated FAM block address is noted in this slot during allocation and a cache hit. The format of the DRAM cache metadata and its retrieval is shown in Fig. \ref{fig:dramcache_metadata}. e.g., when managed as fully associative cache, 16MB cache with 256B block size would require approx. 450KB(64K*7B) of metadata to cover 48-bit physical address space, which is less than 5\% of DRAM cache size. Hence managing DRAM cache metadata is practical.

\subsection{Demand \& Prefetch requests handling mechanism}
\begin{figure}[h]
\centering
\includegraphics[scale=0.23]{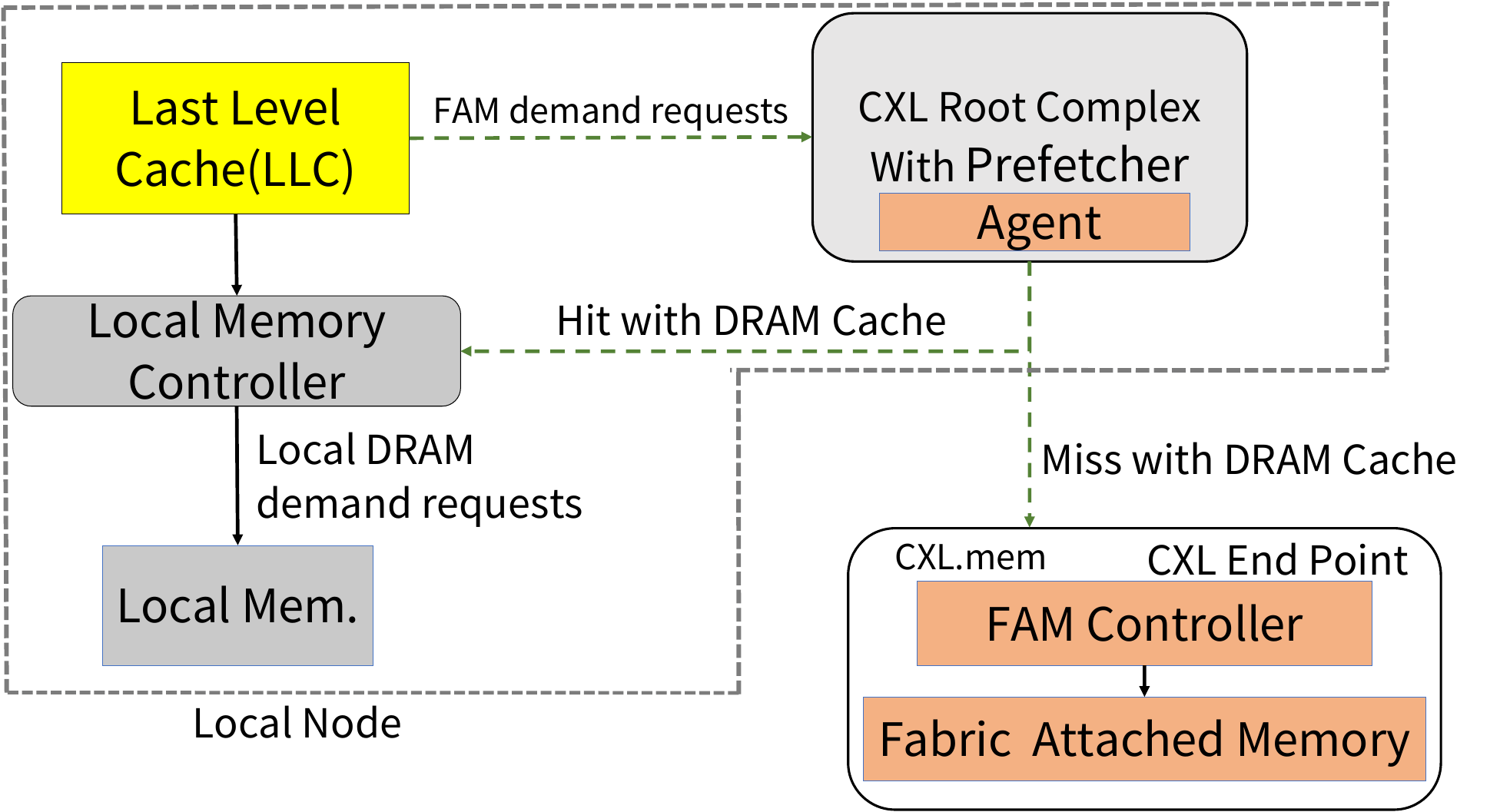}
\caption{FAM demand request flow with prefetcher and DRAM cache}
\label{fig:fam_demand_request_flow}
\vspace{-1em}
\end{figure}

Fig. \ref{fig:fam_demand_request_flow} explains the flow of demand requests with DRAM cache and prefetcher. For every outgoing FAM demand request, the prefetcher is consulted to check if the requested block is present in the DRAM cache. Prefetcher promptly checks the metatdata for the requested block. If the demand block is present(DRAM cache hit), a new request with the DRAM cache block address(obtained from the metadata) is sent to the local memory controller. FAM demand request waits for the response of this new request and will return with response data. As said earlier, this FAM demand request can be a true demand request by application or a core prefetch, either of request type can be served by the DRAM cache. Subsequently the corresponding LRU field in metadata will be updated. If the demand block is not present in the DRAM cache(DRAM cache miss), demand request will proceed per usual to the FAM. 
\vspace{-1em}

Irrespective of the DRAM cache hit status, prefetcher generates prefetch address for every outgoing FAM demand request. Before sending out the prefetch requests, prefetch queue and DRAM cache metadata are checked to see if generated prefetch request is redundant. Prefetch continue to issue stage, once the queue and metadata check clears. When in the issue stage, prefetch request can be dropped if the prefetch queue is full or at pre-defined threshold(eg: 95\%).

Past the issue stage successfully, once the prefetch request's response is received, prefetcher checks the metadata to see if there is any vacancy in the DRAM cache. If there is a vacancy, the prefetch block would be directed to appropriate block of the DRAM cache directly. If there is no vacancy, prefetcher issues an eviction for the LRU block first in DRAM cache and then the corresponding position will be replaced by the incoming prefetched block. 
%Fig. \ref{fig:prefetch_request_flow} shows the flow of prefetch requests.
%In order to minimize the security implications of this caching approach, all the blocks within a 4KB page belong to one requestor. 
%When all the blocks of a page are in DRAM, the page could be migrated from FAM to DRAM.
%\begin{figure}[h]
%    \centering
%    \includegraphics[scale=0.23]{figures/dram_cache_demand_prefetch_flow.pdf}
%    \caption{Request flow. Dotted orange line indicates demand requests flow, Solid black line indicate prefetch request flow.}
%    \label{fig:prefetch_request_flow}
%\vspace{-1.5em}
%\end{figure}

We should note the that meta data cost for DRAM cache management increases linearly with increase in the number of blocks. For a given DRAM cache size, the number of blocks decreases as the block size increases. Hence, an advantage of using larger block sizes for DRAM cache is that the meta-data overhead is decreased. On the flip side, using larger block size can cause increased latencies at FAM.
%We consider block sizes ranging from a cache block of 128 bytes to multiples of cache blocks of up to 1KB. 

\subsection{FAM Controller}
The task of FAM controller is to convert incoming cxl.mem protocol requests to DDR requests that could be ultimately handled by the FAM. In a real system, FAM Controller could adapt a port based design, with each port supporting 8$\times$, 16$\times$, lanes of PCIe/CXL at the frontend, while supporting multiple channels of DDR4/DDR5 memory in backend. Network on chip(NoC) like architectural structures might be present in order to route the incoming requests to appropriate queues that feed into DDR channels \cite{pond}. Vertical scaling of such controllers might be essential to support large number of PCIe lanes/memory channels depending on the pool size and desired number of memory channels.

We abstract the functionality of FAM Controller as components that move incoming requests to DDR channels. All the incoming requests from multiple nodes are filled into the input queue. We assume that FAM controller is aware of the maximum memory bandwidth across all its  supported DDR channels. Hence, the controller scans the input queues at appropriate rate and issues the requests to the respective DDR channel.

With the addition of prefetch to the compute node's root complex. FAM controller now receives two classes of requests - demand and prefetch. In the baseline design, we implement a single input queue, with FIFO scheduling. Requests(both demand and prefetch) from multiple nodes are dispatched to FAM in the order of their arrival. Later, we explain how the demand and prefetch requests can be given different treatment at the FAM through such mechanisms as Weighted Fair Queuing (WFQ).

\subsection{Sub-page block size vs. latency trade-off analysis}
\begin{figure}[h]
    \includegraphics[scale=0.45]{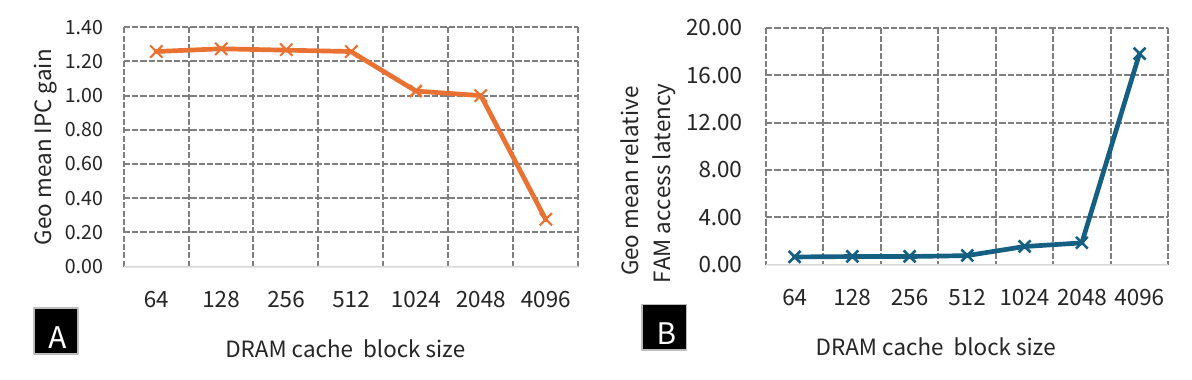}
    \caption{Subfigures A \& B represent geo. mean IPC gain and relative FAM access latency across different prefetch block sizes(Both wrt to baseline)}
     \label{fig:expl_blocksize_analysis}
     \vspace{-1em}
\end{figure}

We performed an exploratory analysis to understand the  block size vs. latency tradeoff for sub-page block DRAM cache prefetching. We observe the IPC and average FAM access latency by varying the DRAM cache block size. Fig. \ref{fig:expl_blocksize_analysis} show our analysis.

As DRAM cache block size increases from 64-512B, IPC gain stays mostly constant with marginal improvements for 128B, 256B block sizes. Beyond 512B, average IPC gain decreases due to increase in relative FAM latency. Plus, moving a FAM page on touch to DRAM cache (4096B block size) observes about 17 $\times$ increase in relative FAM latency, resulting in substantial IPC decrement. Based on this analysis, we consider 128, 256, 512 as block sizes for DRAM cache. Using multiple of CPU cache block size(64B) for DRAM cache, will amortize the delays due to flit packaging/serialization at fabric, as well as reduce the hardware overhead of meta-data management.

\section{Prefetch Optimizations}
Our optimizations to the DRAM cache prefetcher are aimed at mitigating the interference between demand and prefetch requests, there by enhancing the utility of FAM accesses. To achieve this we propose two approaches - Weighted Fair Queueing at memory node, Prefetch Bandwidth adaptation at compute node. We describe the design and implementation of both of these optimizations below.

\subsection{Weighted Fair Queueing(WFQ)}
We consider WFQ as a generic means for providing priority to demand requests over prefetches at the FAM. By giving a higher weight to demand requests, we can provide a higher slice of FAM bandwidth to demand requests. Under congestion, WFQ ensures demands are served with priority over prefetches, there by pontentially mitigating queueing delays for demands due to prefetches.

%The main goal of using WFQ at the memory node is to prioritize demand requests over prefetch requests during the times of heavy traffic to the FAM. This design choice is valid because the processors at the compute node are waiting for the completion of demand requests whereas the same thing doesn't always hold true with the prefetch requests.

We enhance the baseline queueing implementation of the FAM controller, by replacing the single input queue with two input queues, one each for demand and prefetch. Double queue implementation enables us to issue prefetches and demands at independent rates. We take advantage of the prefetch request tagging by the prefetcher at the compute node, to identify the placement of the incoming requests into their respective queues. Both core prefetches and DRAM cache prefetches are placed in the prefetch queue.

We use a WFQ scheduler to issue requests from the two queues to the FAM. We use work-conserving deficit weighted round-robin(DWRR)\cite{dwrr} algorithm to select the queue from which the request should be issued. W-weight is used to indicate how much weightage demand requests are given relative to the prefetch requests. Pseudo code of the our algorithm is as shown in Alg.\ref{alg:demand_prefetch_issue}

\begin{algorithm}
    \caption{Demand/Prefetch Issue Algorithm}
    \SetAlgoLined
    \SetKwProg{Fn}{Function}{:}{}
    \SetKwFunction{IssueRequests}{IssueRequests}
    \SetKwFunction{IssueDemandRequests}{IssueDemandRequests}
    \SetKwFunction{IssuePrefetchRequests}{IssuePrefetchRequests}
    \Fn{IssueRequests()}{
    current\_round += (current\_weight+1)\%(W+1)\;
    demand\_queue\_status = CheckDemandQueue()\;
    prefetch\_queue\_status = CheckPrefetchQueue()\;
    r = prefetch\_block\_size/demand\_block\_size\; 
    \eIf{current\_round != 0}{
    \uIf{demand\_deficit $<$ max\_demand\_deficit}{
            demand\_deficit += quantum\;
        }
        \uIf{demand\_queue\_status and demand\_deficit$>$0}{
          IssueDemandRequests()\;
          demand\_deficit = demand\_deficit-1\;
          
        }
        \uElseIf{prefetch\_queue\_status and prefetch\_deficit $>$ r }{
            IssuePrefetchRequests()\;
            prefetch\_deficit = prefetch\_deficit-r
        }
      }{
      \uIf{prefetch\_deficit $<$ max\_prefetch\_deficit}{
      prefetch\_deficit += quantum
      }
      \uIf{prefetch\_queue\_status and prefetch\_deficit$>$r}{
          IssuePrefetchRequests()\;
          prefetch\_deficit = prefetch\_deficit-r\;
          
        }
        \uElseIf{demand\_queue\_status and demand\_deficit $>$ 0 }{
            IssueDemandRequests()\;
            demand\_deficit = demand\_deficit-1
        }
      }
    }
    \label{alg:demand_prefetch_issue}
\end{algorithm}
For every issue cycle, the IssueRequests() function is called to see either of demand or prefetch requests can be issued. We use the current\_round variable to track the round number within a W+1 round window, when the prefetches and demands are serverd in 1:W ratio. Prefetch requests are preferred in only one round of the window. During the rest of rounds, we prefer to issue demand requests. Due to the scheduler being work conserving in nature, if the preferred choice of requests are not available, we try to issue the other type of requests.
When its the demand turn, if the demand\_deficit did not exceed maximum permissible deficit(max\_demand\_deficit), we increment the demand\_deficit. If the demand queue is non-empty and demand\_deficit is greater than 0, then we issue a demand request, resulting in decrement of the demand\_deficit by 1. If either demand queue is empty or demand does not have enough deficit, we try to issue prefetch requests, which is again subject to prefetch queue non-emptiness and status of the prefetch deficit. The reverse applies for issue logic when it is prefetch's turn.

%IssueDemandRequests() removes a request from the head of the demand queue and issues it to the FAM. Similarly functionality is implemented by the IssuePrefetchRequests() procedure for the prefetch queue.
In order to account for the difference in prefetch and demand block sizes. We enforce that prefetch\_deficit needs to be at least the ratio of prefetch block size and demand block size, for a prefetch to be issued to the FAM. Our proposed DRAM prefetcher works in conjunction with core prefetcher. So, WFQ need to handle both CPU cache block core prefetches as well as sub-page block DRAM cache prefetches. In our implementation, when its the prefetch turn, based on the available deficit we issue either a core prefetch or DRAM cache prefetch. Block size is taken into account when updating deficit post issue.

\subsection{Prefetch Bandwidth Adaptation}
\label{BLAP_bandwidth_adaptation}
%While the look ahead distance adaptation tries to optimize the timeliness of prefetch requests based on the latency of demand and prefetch requests, it doesn't adaptively change the number of prefetch requests that are issued to the FAM.
In the baseline prefetcher, we generate a fixed number of prefetch requests(prefetch degree) for every LLC miss and issue those requests depending on the prefetch queue availability. When the FAM device is saturated due to high number of demand requests, issuing prefetch requests would increase demand latency and could hurt performance if the demand and prefetch requests are queued in a single queue. Under such conditions, it is better to dial back the prefetch request issue rate and wait till the FAM has enough bandwidth to accommodate prefetch requests.
%Baseline prefetcher design lacks this kind of system-level feedback. 
Hence to incorporate such feeedback, we implement prefetch bandwidth adaptation at the source. 

%\textcolor{red}{This needs work. This is too much detail. We could just say "We measure the average demand latency. The average demand latency is employed to adapt the prefetch issue rate. When demand latency is above 150\% of minimum demand latency, the prefetch issue rate is throttled. The amount of reduction of prefetch rate is a function of how far the average demand latency is from the minimum....."}
We take a sampling based approach, to adapt the prefetch issue rate. To learn about the system state, we add event counters to the root complex's prefetcher state. Each counter stores two values, instantaneous value and average value. The instantaneous values of counters are scanned and reset during each start of each sampling cycle. Average value of the event counter stores the exponential moving average of the respective instantaneous values. 
The descriptions of event counters that are stored in prefetch state are as shown in Table.\ref{tab:event_counter_table}.
\begin{table}
    \centering
    \begin{tabular}{|c|c|}
        \hline
        \textbf{Event counter} & \textbf{Description} \\ 
        \hline  
         \texttt{demand\_requests\_issued}  & demand requests \\
                                            & issued to the FAM \\ \hline
         \texttt{demand\_requests\_returned} & demand requests \\
                                             & return from FAM \\ \hline
         \texttt{demand\_requests\_total} & Total demand requests \\
                                          & arrive at the prefetcher \\\hline
         \texttt{prefetch\_requests\_issued} & prefetch requests \\
                                             & issued to FAM \\ \hline
         \texttt{minimum\_demand\_latency} & minimum demand \\
                                           & read latency in recent history. \\ \hline
    \end{tabular}
    \caption{Description of event counters}
    \label{tab:event_counter_table}
    \vspace{-2.5em}
\end{table}
\vspace{-1em}
\begin{figure}[h]
    \centering
    \includegraphics[scale=0.2]{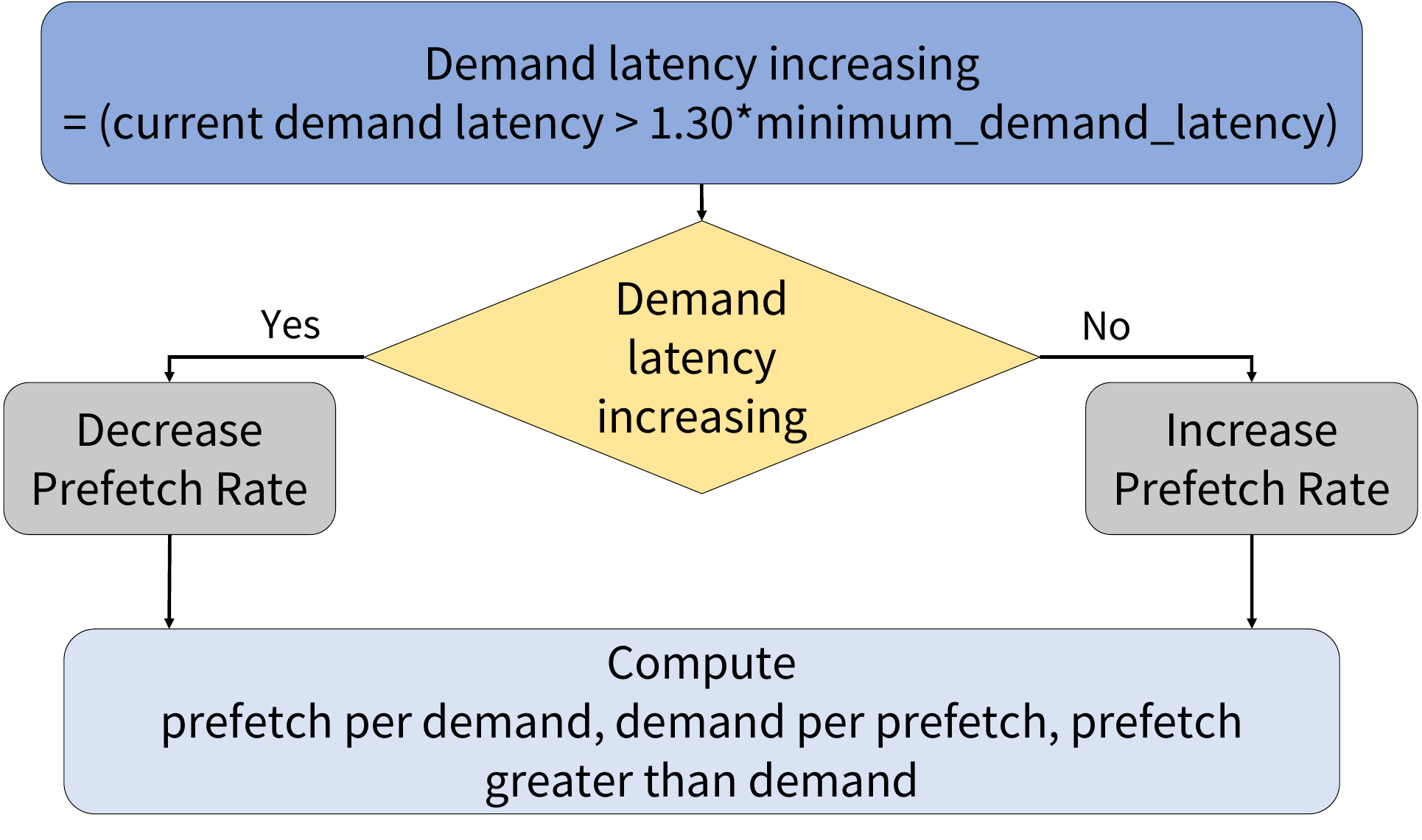}
    \caption{State diagram for prefetch BW adaptation algorithm}
    \label{fig:bw_algorithm}
    \vspace{-1em}
\end{figure}

Abstract-level logical flowchart of our bandwidth adaptation algorithm is shown in Fig.\ref{fig:bw_algorithm}. The algorithm is executed every sampling cycle. Our key idea here is to reduce the prefetch issue rate when FAM is experiencing congestion. 
%The congestion is likely caused due to excess prefetch requests. 
Hence, we track demand request latency and decrement the prefetch issue rate whenever the latency start growing. Measuring demand latency, we compare the measured demand read latency with minimum achievable demand read latency. Minimum achievable demand read latency is unknown,dynamically changing, and depends on fabric topology. We approximate minimum achievable demand read latency to lowest average value in the recent past. By closely tuning the past history, one can tweak the agility of prefetch throttling. If the latency is above 125\% of minimum demand read latency(above the noise level), it might be because of congestion at FAM, then we proceed to decrease the prefetch issue rate. Else, we can increase the prefetch issue rate.
%With this approach, we can adaptively increase/decrease the prefetch request rate without the knowledge of bandwidth. }

%\textcolor{red}{Are we using AIMD or MIMD now? We need to rewrite this based on our current algorithm.}

%To decrease and increase, the prefetch rate, we use multiplicative increase and decrease(MIMD) approach. 
We employ Multiplicative Increase and Multiplicative Decrease (MIMD) \cite{Jain-congestion} for adjusting the prefetch rate. In our implementation, we set the increase factor to 1.125(12.5\% over prev. value). The decrease factor is determined dynamically based on the observed behavior. The decrease factor is a function of prefetcher accuracy, with higher accuracy resulting in slower decreases. We expect this to result in more accurate prefetches to be issued when multiple applications are competing for bandwidth at FAM. In addition, we mimic RED \cite{RED, PERT} at the source and make the decrease factor linearly dependent on the difference of observed latency and minimum read latency, when the latency is above the threshold of 125\% of minimum latency. 25\% is heuristic, we chose for the noise level.

While both WFQ and bandwidth adaptation at the source are trying to address the same problem, we evaluated both the schemes to evaluate their relative merits. CXL memory nodes could provide additional functionality and our evaluation of WFQ is intended to understand the implications of augmenting a memory node with WFQ. We compare the two approaches to throttling prefetches.

\begin{table}[h]
    \centering
    \begin{tabular}{|c|c|}
    \hline
        Processor &  8 out-of-order cores \\
                  & clock: 3.3 GHz, 6 issue/cycle \\ 
                  & max pending transactions : 16 \\ \hline
        L1 cache & 32 KB, 4 ways \\
         & 4 cycle access latency\\ \hline
        L2 cache & 256 kB, 8 ways \\
        & 12 cycle access latency \\ \hline
        L2  &  Signature Path \\ 
        Cache Prefetcher & Prefetcher(SPP) \cite{spp}\\ \hline
        L3 cache & 8 MB, 16 ways \\
        & 30 cycle access latency \\ \hline
        Local memory & DDR4-3200\\
        & 2 channels, 2 ranks \\ \hline %TODO: Add DRAM timings from the config file
        Nodes & 1-4 \\ \hline
        CXL Network &  256B flit-size, Min-packet size: 28B \\
        & Bandwidth: 128 GB/s/direction\\ 
        & Min. Latency: 70ns  \\ \hline
        Per-Node & 256\\
        prefetch queue size & \\ \hline
        Pooled FAM & DDR4-2400 \\
        & 2 channels, 2 ranks\\ \hline
    \end{tabular}
    \caption{Simulated system configuration}
    \label{tab:sys_configuration}
    \vspace{-3em}
\end{table}
\hfill
\begin{table}[h]
   \centering
    \begin{tabular}{|c|c|c|c|}
    \hline
    \textbf{Benchmark} & \textbf{Workload} & \textbf{FAM usage} \\
    \textbf{suite} &  & \\ \hline
    SPEC17  & 603.bwaves\_s & 824 MB \\ 
            & 607.cactuBSSN\_s & 257 MB\\ 
            & 619.lbm\_s &  1.55 GB\\
            & 628.pop2\_s &  590 MB\\
            & 649.fotonik3d\_s & 587 MB\\ 
            & 654.roms\_s & 245 MB\\ 
            & 657.xz\_s & 561 MB \\ \hline
    Splash 3 & LU & 515 MB\\ 
            & FFT & 625 MB\\  \hline
    GAP & bfs &  864 MB\\ 
        & cc &  802 MB\\ 
        & bc & 593 MB\\ 
        & sssp & 545 MB\\ \hline
    PARSEC & dedup & 868 MB\\
           & facesim & 188 MB\\
           & canneal & 849 MB\\ \hline
    NPB & mg & 431 MB \\
        & is & 1 GB\\ \hline
    XSBench & XSBench & 611 MB\\ \hline
    \end{tabular}
    \caption{Benchmark configurations}
    \label{tab:benchmark_configuration}
    \vspace{-3em}
\end{table}

\section{Evaluation}
\subsection{Methodology}
%TODO: Following citations are missing - PIN, Ariel -
% TODO: Mention that we used DDR5 based timing measurements for local memory and NVM performance characterstics for the pooled memory
We evaluate DRAM cache prefetcher along with optimizations using SST\cite{sst} simulation components. 
%SST provides infrastructure for cycle accurate simulation of several computer system components like processor, memory, interconnect network etc. 
We used Ariel, a pin-tool based processor front-end simulator, to simulate compute nodes. 
%Ariel instruments the instructions of the given benchmark, generating memory read/write events that are subsequently pushed into memory hierarchy simulation. 
Ramulator\cite{ramulator} was to model both local memory(DRAM) and FAM devices .
%NVDIMM model from Messier\cite{messier} was used to simulate the pooled memory. 
%Samba\cite{samba} was used to model the Memory management Unit(MMU) \& Translation Lookaside Buffer(TLB). 
We used Opal\cite{opal} to emulate the operating system's memory allocator and page fault handler. 
%Memory allocation is simulated during first access to an unmapped address. 
Opal allows us to configure the memory footoprint between local DRAM and pooled FAM. CXL network is simulated by provided flit based network model, with programmed delay and bandwidth.
%Our MMU treats mmap calls and first access to unmapped address as page faults, invoking page fault handler of the opal component. Opal allows for the configurable memory allocation policy for benchmarks. For our evaluation, we used static proportional policy, where local and FAM will share the total benchmark pages in the ratio of their capacities(1:8 if not mentioned otherwise). 

%Recent CXL implementation\cite{intel_cxl_implementation} by Intel demonstrated that the CXL port latency(round trip) to be around 25ns. We assume that there exists 3 such ports on the path from each node's LLC to FAM device, hence we model the CXL network latency to be 75ns.  We employ prefetch block size of 256B. We found 256B to be a good tradeoff between exploiting spatial locality through larger prefetches and reducing waiting times behind larger prefetches. This also matched the CXL flit size of 256B. 

We evaluated 19 memory intensive workloads from  benchmark suites like SPEC\cite{spec17}, PARSEC\cite{parsec3}, GAP\cite{gapbs}, Splash3\cite{splash3}, and NPB\cite{npb}. Modern servers contain 64-128 processors per node equipped with 100'GB of main-memory. Simulating such a system, is impractical given the simulation speeds. For realistic simulation schedules, we simulate a system with scaled down configuration, that runs regions of interest(ROI) within each benchmark. We expect that our simulator and the corresponding performance characterstics to scale to a larger configuration, with no issues. . The simulated system configuration is detailed in Table.\ref{tab:sys_configuration}. Evaluated applications and their respective memory footprints are shown  in Table \ref{tab:benchmark_configuration}.

We have simulated both single and multi node systems accessing FAM memory pool. For multi-node systems, we ran copies of same application on different nodes, as well as different applications on different nodes. In multinode systems, we expect the higher loads at FAM to result in tighter availability of bandwidth and hence possibly higher congestion. We evaluated 7 workload mixes for 4-node system. 

Below we define figures of merit, terms, and configurations that we use in discussion through the rest of the section
\begin{enumerate}
    \item \textit{Core pretcher -} Each of the node in our system, comprises a multi-core CPU. Within each core, a prefetcher is at the L2 cache level. 
    \item \textit{Baseline configuration -} Workload running with no core prefetching and DRAM cache prefetching enabled. Core prefetcher is turned ON for all config, except in baseline.
    \item \textit{all-local configuration -} Workload running with core prefetching, with entirety of its memory footprint residing in local DRAM.
    \item \textit{allocation ratio(X) -}  Workload's memory footprint divided between FAM and DRAM in ratio of X:1 respectively. 
    \item \textit{IPC gain -} Ratio of IPC for a given workload config to that of workload in baseline config .(Higher the better)
    \item \textit{Relative FAM latency -} Ratio of the average FAM access latency for a workload in a given configuration to that of workload running in baseline configuration.(Lower the better)
    \item \textit{Relative DRAM prefetch requests issued - } Ratio of DRAM cache prefetches issued for a given config to that of DRAM cache prefetches issued with no optimizations(FIFO scheduling and no prefetch BW adaptation).
    \item \textit{Demand hit fraction - }Fraction of demand requests that miss LLC, that hit in DRAM cache.
    \item \textit{Core Prefetch hit fraction - }Fraction of core prefetch requests that miss in LLC, that hit in DRAM cache. 
\end{enumerate}
%\textcolor{red}{
%Through our evaluation, we desire to analyze the following aspects of our proposal}
%\begin{itemize}
%    \item Performance gain due to Prefetch Bandwidth Adaptation at compute node $\S$\ref{subsec:bw_adapt}
%    \item Performance gain due to Weighted Fair Queueing(WFQ) at memory node $\S$\ref{subsec:wfq_scheduling}
%    \item DRAM cache prefetcher behaviour in multi-node system running different applications
%    \item Senstivity of our design to different system parameters
%        \begin{enumerate}
%            \item Workload memory footprint division ratio-FAM/DRAM capacity ratio
%            \item DRAM cache size
%            \item Network latency
%        \end{enumerate}
%\end{itemize}
%}
\subsection{Performance gain with Prefetch Bandwidth Adaptation}
\label{subsec:bw_adapt}
For this analysis, we run workloads in 1,2, and 4 node system configurations. Each workload ran in 3 prefetch configurations - core prefetcher turned ON, core prefetcher+DRAM cache prefetcher turned ON, core prefetcher+DRAM cache prefetcher+prefetch BW adaptation turned ON. We call the second configuration, non-adaptive DRAM prefetch.  
%We measure IPC gain wrt baseline, Relative FAM latency, Relative DRAM cache prefetches issued, demand hit fraction and core prefetch hit fraction for each experiment. 
Each workload has memory allocation ratio of 8.
\begin{figure}
\centering
\includegraphics[scale=0.45]{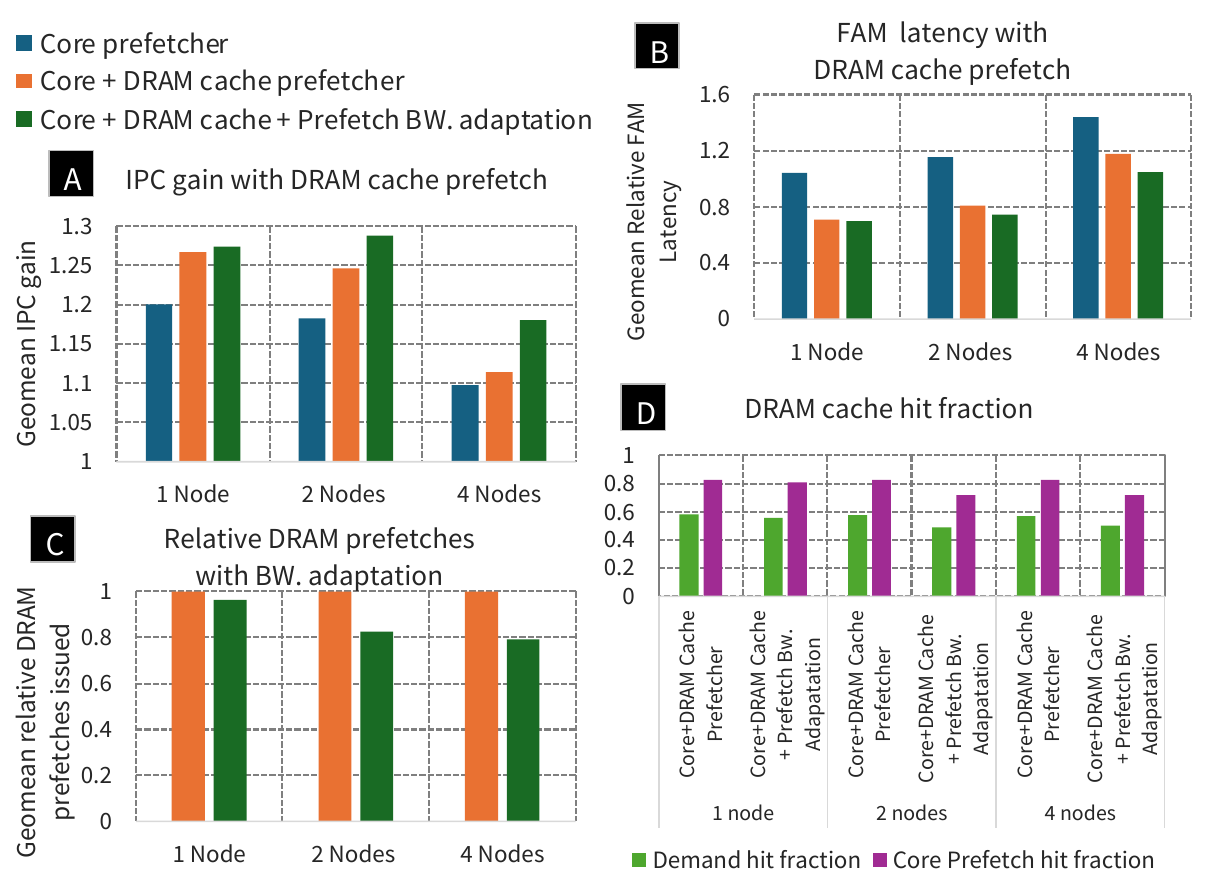}
\caption{Evaluation of DRAM cache prefetcher with prefetch bandwidth adaptation}
\label{fig:bw_adapt_summary}
\vspace{-1.5em}
\end{figure}

Fig. \ref{fig:bw_adapt_summary} outlines the results of our experimentation. \ref{fig:bw_adapt_summary}A  shows geomean IPC gain of all benchmarks, for each prefetch configuration, across 3 node configuration. Across the board, DRAM prefetching improves overall performance compared to core prefetching. Core prefetching resulted in IPC gain of 1.20, 1.18, 1.10 for 1,2,4 node systems. With both core prefetching and DRAM cache prefetching turned ON, the same IPC gains increased to 1.26, 1.24, 1.11 respectively. 
%\textit{Hence, DRAM cache prefetching(non-adaptive) resulted in 6\% IPC improvement for 1,2 node systems over core-prefetching}. 
Performance improvement comes from reduction in FAM access latency, as indicated in Fig. \ref{fig:bw_adapt_summary}B. DRAM cache prefetching reduced average FAM access latency by 29\% and 34\% for 1,2 node systems respectively.

Prefetch BW adaptation further enhanced the performance of DRAM cache prefetching, for 2 and 4 node systems. \textit{BW adaptation resulted in 4\% and 8\% IPC improvement over non-adaptive DRAM cache prefetching for 2 and 4 node systems.} Non-adaptive DRAM cache prefetching performed poorly in 4 node system configuration, with no IPC improvement over core prefetching, thus emphasising the importance of BW adaptation in bandwidth constrained systems. BW adaptation resulted in decrement of 7\% and 13\% in relative FAM latency  over non-adaptive DRAM prefetch.

Fig. \ref{fig:bw_adapt_summary}C presents the relative no. of DRAM prefetch requests. Results indicate that adaptation caused 18\% and 21\% less DRAM cache prefetches to be issued to FAM, in 2-node, 4-node systems respectively. Decreased DRAM prefetch issue rate resulted in decreased demand and core-prefetch hit rate, according to analysis presented in Fig. \ref{fig:bw_adapt_summary}D. For instance, BW adaption reduced the demand and core-prefetch hit fraction from 57\% and 83\% to 50\% and 72\%, for 4-node system. Performance improved despite hit fraction decrement, which reveals that prefetch requests are causing considerable queuing delays at FAM.
\begin{figure}
    \centering
    \includegraphics[scale=0.42]{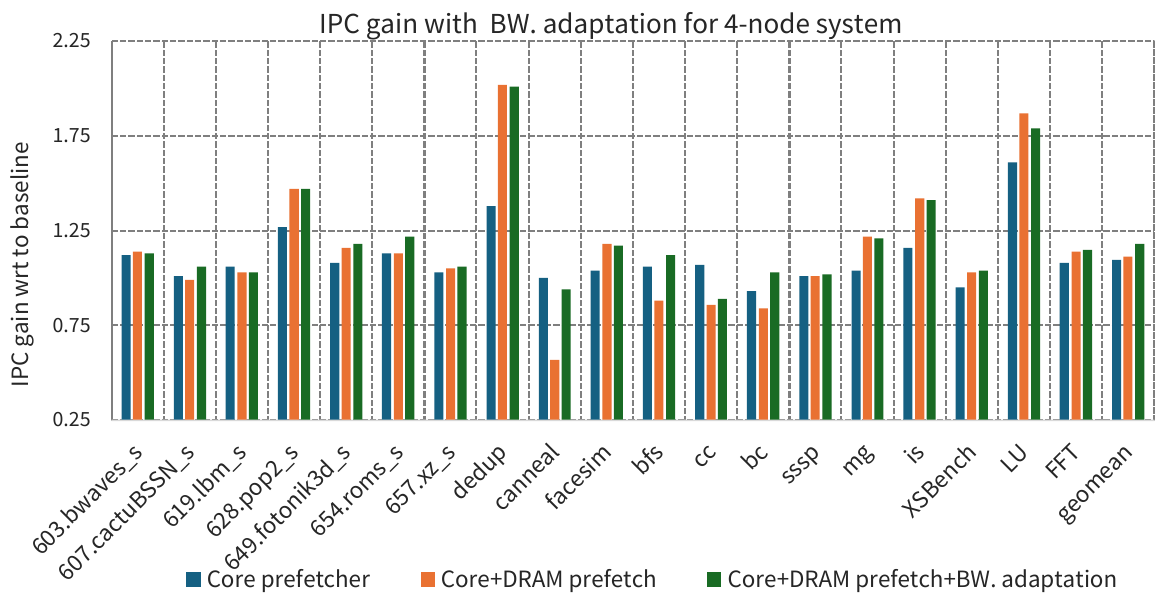}
    \caption{IPC gain due to BW. adaptation for 4-node system. Geomean from this analysis is represented in "4 Nodes" in Fig. \ref{fig:bw_adapt_summary}A}
    \label{fig:4node_bw_adaptation}
    \vspace{-2em}
\end{figure}

Further, we analyze the IPC gain with prefetch bandwidth adaptation for 4-node system across different benchmarks, as shown in Fig. \ref{fig:4node_bw_adaptation}. DRAM cache prefetch significant improves IPC for applications like dedup, LU, 628.pop2\_s, mg, is, and facesim. Workloads like canneal, bfs, cc, and bc saw IPC decrement with DRAM cache prefetch, possibly due to increased FAM latency. BW adaptation was able to improve the IPC substantially for these applications, expect for cc.

BW adaptation would mitigate congestion only when the DRAM prefetches are in some proportion responsible for creating it. This is because our algorithm implementation throttles only DRAM cache prefetch issue rate. BW adaptation would be of little help if core prefetches are responsible for congestion. Future implementations of our prefetch throttling algorithm can relay the congestion occurrence to the CPU cache controllers, enabling dynamic throttling of CPU prefetch requests.

\subsection{Performance gain with WFQ scheduling}
\label{subsec:wfq_eval}
\begin{figure}
    \centering
    \includegraphics[scale=0.45]{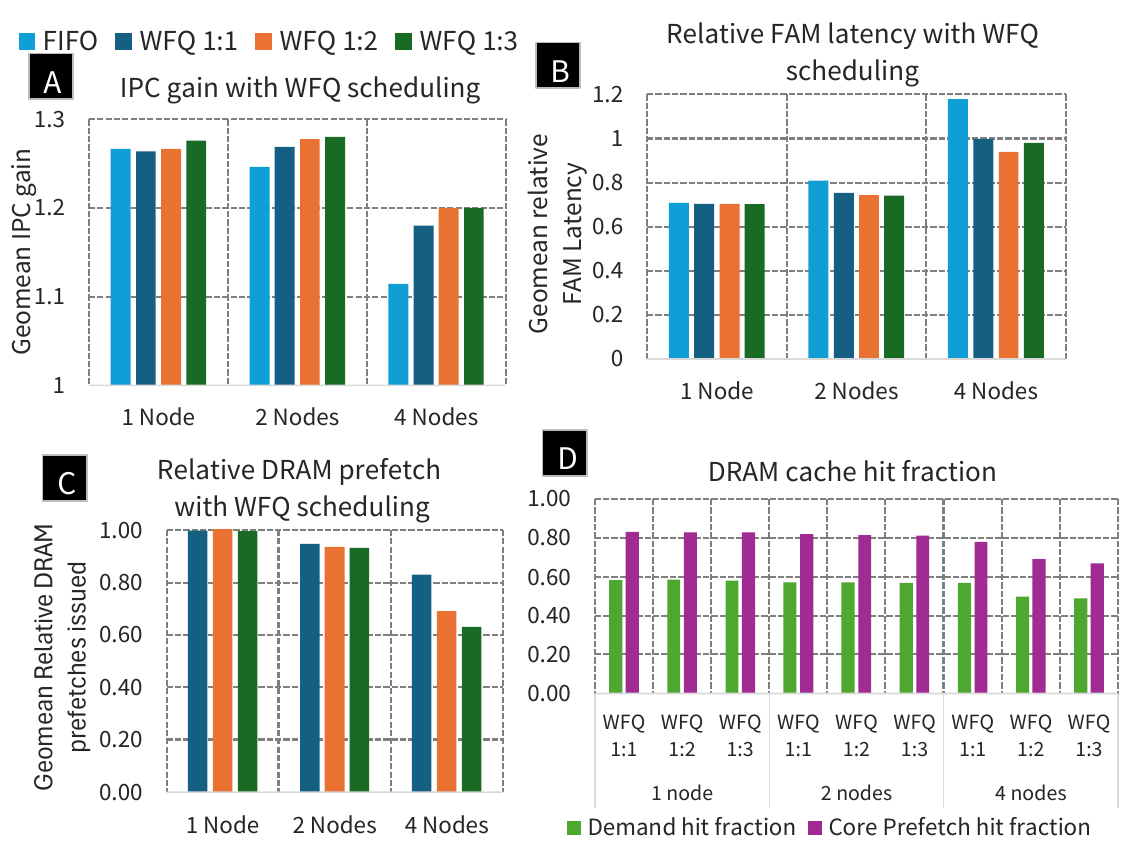}
    \caption{Evaluation of DRAM cache prefetcher with WFQ scheduling}
    \label{fig:wfq_summary}
    \label{subsec:wfq_scheduling}
    \vspace{-1em}
\end{figure}
We evaluate our WFQ scheduling algorithm with 3 weights-1,2,3 (weight of 3 indicates demands and prefetches are served in 3:1 ratio). Each workload is run in 1,2,4 node system configuration, with WFQ scheduling at the FAM controller, and with different weights. 
%We collect the same metrics as in $\S$\ref{subsec:bw_adapt} for our analysis. 
Performance of WFQ with different weights is compared to FIFO scheduling(non-adaptive) prefetch. Core prefetcher is active for all the configurations examined here. Fig. \ref{fig:wfq_summary} shows our results.

Geomean of IPC across all benchmarks for WFQ algorithm for a given weight and node configuration is shown in Fig. \ref{fig:wfq_summary}A. WFQ improves the performance over fifo scheduler for 2,4 node systems. \textit{Weights 1,2,3 improve the average IPC by 8\%(3\%), 9\%(4\%), 9\%(4\%) over FIFO scheduler for a 4(2) node system.}
%Weights 1,2,3 improve the average IPC by 8\%, 9\%, 9\% over FIFO schedulerfor 4 node system.
Again, the increase in IPC is due to decrement in relative FAM latency. For a 4(2) node system, average relative FAM latency is reduced by 24\%(10\%). 
%For 4 node system, average relative FAM latency is reduced by around 24\%. 
Given that BW adaptation resulted in 7\% IPC improvement over FIFO scheduler, WFQ marginally performs better.

WFQ also resulted in less number of DRAM prefetches to be issued. For a 4 node system, WFQ with weights 1,2,3 resulted in 17\%, 31\%, 37\% decrement in average relative DRAM prefetches issued. Such behavior is expected because, as the weightage to demands increase, prefetch request latency increases, filling the prefetch queue, subsequently causing less number of prefetches to be issued. Fig. \ref{fig:wfq_summary}D shows the demand and core-prefetch hit fraction with WFQ across different node configurations and weights.
\begin{figure}
    \centering
    \includegraphics[scale=0.45]{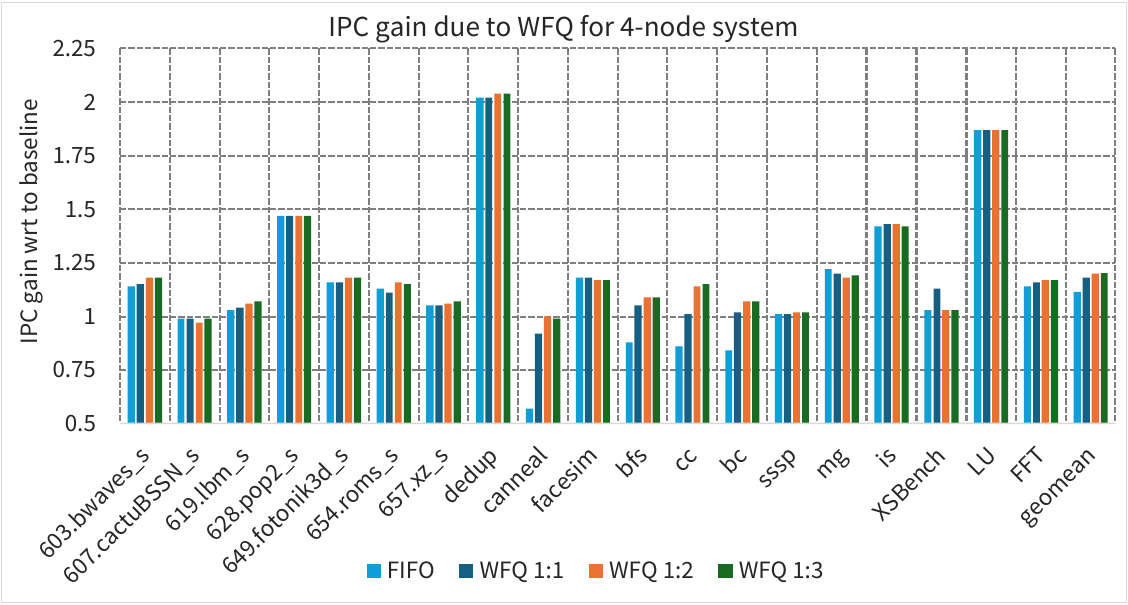}
    \caption{IPC gain due to WFQ for 4-node system. Geomean from this analysis is represented in "4 Nodes" in Fig. \ref{fig:wfq_summary}A}
    \label{fig:4node_wfq}
    \vspace{-2em}
\end{figure}

Additionally, we analyze the IPC gain with WFQ for 4-node system across different benchmarks, as shown in Fig. \ref{fig:4node_wfq}. Set of workloads that benefited from BW. adaptation, benefited from WFQ as well. Interestingly, cc application see IPC improvement with WFQ but not with BW. adaptation.
Due to placement of both core prefetch and DRAM cache prefetch requests into the same queue. WFQ can potentially mitigate congestion due to either of the request type. Due to this reason, WFQ performs marginally better in comparison to bandwidth adaptation.
\subsection{Performance analysis of multi-node workload mixes}
\begin{figure}
    \centering
    \includegraphics[scale=0.5]{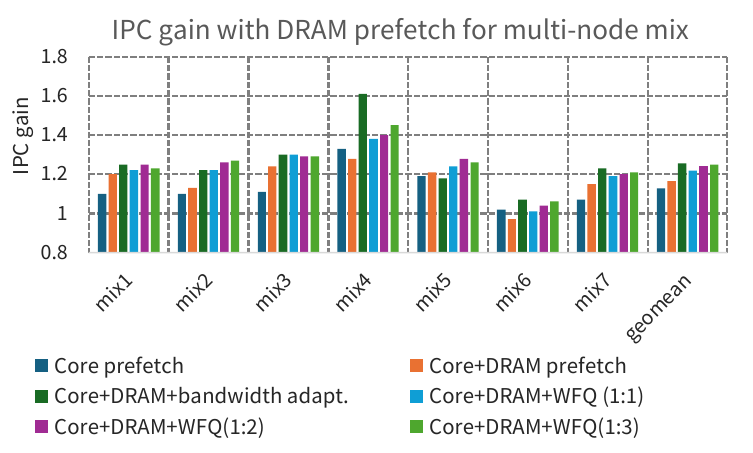}
    \caption{Performance gain with different configurations of DRAM prefetch across 7 multi-node workload mixes}
    \label{fig:multinode_mix_ipc}
    \vspace{-1em}
\end{figure}
\label{subsec:mixes}
Combining the methodology of $\S$\ref{subsec:bw_adapt} and $\S$\ref{subsec:wfq_eval}, we evaluated our 7 multi-workload mixes with a total of 5 prefetch configurations. Each mix is run on a 4 node system. Fig. \ref{fig:multinode_mix_ipc} shows the IPC gain for each of the workload mix. BW adaptation and WFQ provide equivalent IPC improvement for mix1, mix3, mix6 and mix7. Mix5 saw slight IPC decrement with BW adaptation, but performance improvement with WFQ. BW adaptation outperformed WFQ by 16\% for mix4. WFQ outperformed BW adaptation by 5\% for mix2. On an average, BW adaptation and WFQ resulted in 10\% and 9\% IPC over non-adaptive prefetcher(FIFO scheduler) respectively.

This analysis reveals to us that both of these approaches are useful in resolving congestion at FAM. But the relative performance gain due to either of these techniques depends not just on the workload alone, but also on co-existing workloads, that are accessing FAM.  
\subsection{Performance improvement across allocation ratio's}
\begin{figure}
    \centering
    \includegraphics[scale=0.55]{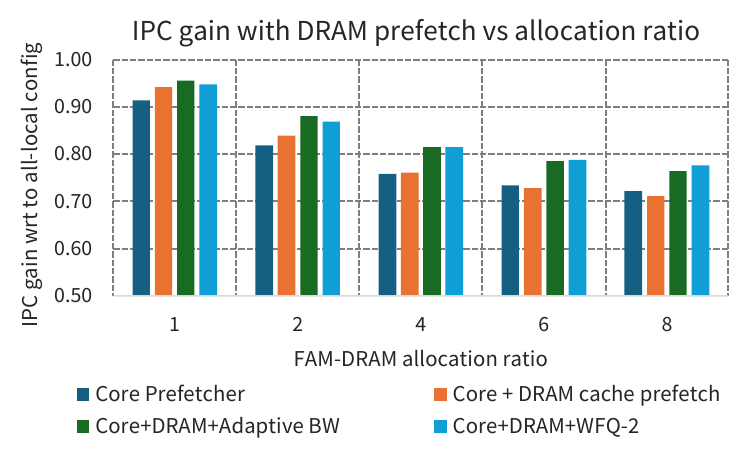}
    \caption{IPC gain wrt to performance with entire memory in local DRAM,}
    \label{fig:fam_dram_analysis}
    \vspace{-1em}
\end{figure}
We analyze the impact of DRAM cache prefetch along with proposed optimizations across different allocation ratios. We considered 4 prefetch configurations for this experiment - Core prefetcher ON, Core+DRAM cache prefetch, Core+DRAM+BW. adaptation, Core+DRAM+WFQ(1:2). We vary the allocation ratio 1 to 8 and measure the IPC gain with respect to all-local configuration for each benchmark, for a 4-node system. Fig. \ref{fig:fam_dram_analysis} shows the geo. mean of IPC gains of all benchmarks running with a given allocation ratio and prefetcher configuration. 

Firstly this analysis reveals that utility of core prefetching decreases as FAM usage increases. With allocation ratio of 1, workloads saw an average of 10\% IPC decrement, but with allocation ratio of 8, workloads saw an average of 28\% IPC decrement. DRAM prefetch helps the bridge the performance gap between pooled memory configuration and all-local configuration, improving the IPC by an average of 5\%-6\% across all the allocation ratios. Importance of BW. adaptation and WFQ are more evident in higher allocation ratios, non-optimal DRAM cache prefetching result in no IPC improvement for 4, 6, 8 allocation ratios. 
\subsection{Sensitivity to DRAM Cache Size}
\begin{figure}
    \centering
    \includegraphics[scale=0.45]{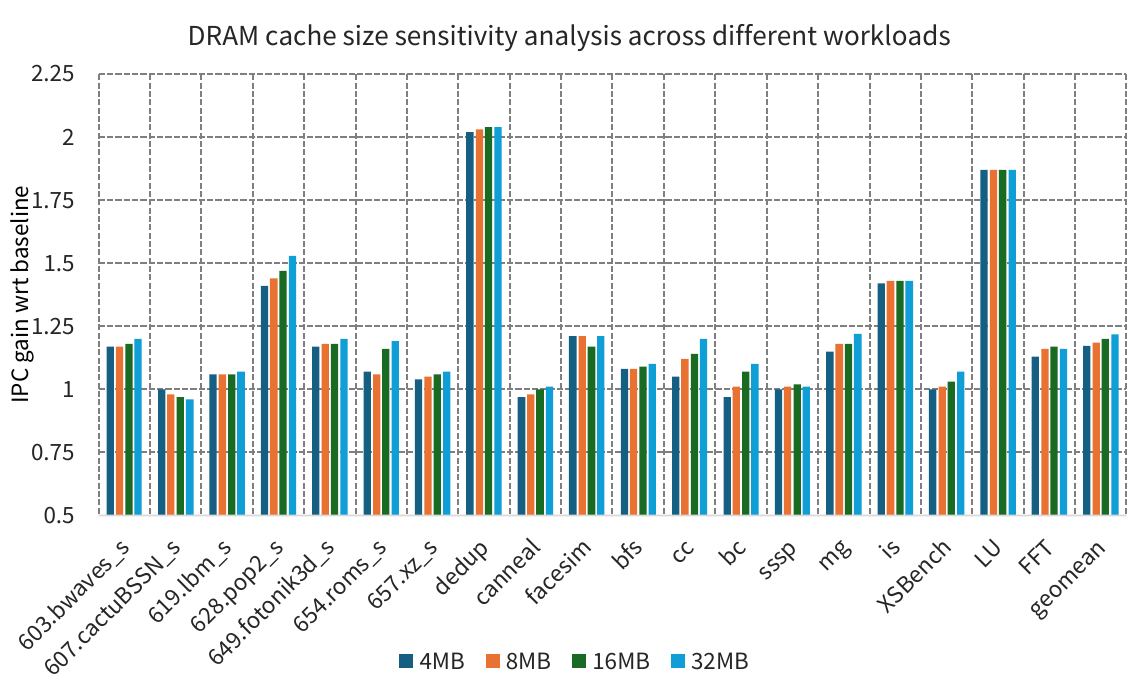}
    \caption{DRAM cache size sensitivity analysis, with DRAM cache sizes varied from 4-32MB}
    \label{fig:dram_cache_sensitivity}
\end{figure}

To analyse the sensitivity of our design to DRAM cache size, we simulated a 4-node system, running same copies of the given workload, with varying DRAM cache sizes. To potentially negate the effect of contention, we have used WFQ scheduling policy with weight as 2. Our analysis is shown in Fig. \ref{fig:dram_cache_sensitivity} shows the results of our analysis. Benchmarks like 628.pop2\_s, 654.roms\_s, cc, bc, XSBench showed sensitivity to DRAM cache size, their respective IPC gains increased with increase in DRAM cache size. On average, DRAM cache size of 4MB, 8MB, 16MB, 32MB resulted in  IPC gain of 1.17, 1.19, 1.20, 1.22. IPC improved by 5\% as the DRAM cache size increased from 8 to 32MB on average.
\section{Related Work}
Earlier work has considered feedback to control the prefetching rate from DRAM into LLC \cite{fdp}.Our work differs from this work in significant ways, considers prefetching from FAM into DRAM, employs more sophisticated prefetchers and employs separate queues and priorities for demand and prefetch requests. Separate queues for prefetch and demand requests and fair queuing have been studied earlier in the context of multiprocessor systems where resources are shared \cite{friendly-fire,prefetch-shared}. We leverage this earlier work in a different context. 

Blue \cite{timely-prefetcher} 
%and APT-GET \cite{apt-get} 
considers timeliness of prefetches in order to issue prefetches sufficiently ahead. However, the timeliness measures in earlier work do not consider dynamic latencies at the memory system. Our approach here also separates prefetches and demands into separate queues. Criticality aware prefetchers consider the criticality of prefetches in reducing stalls \cite{CLIP}. Our system here operates outside the processor chip and doesn't have access to such information. 

Prefetching from remote memories has received significant recent attention \cite{span,infiniswap, leap,fastswap}, in order to reduce the cost of moving pages from one node to another while allowing memory to be shared across nodes on a network. Our work takes motivation from this work, but considers hardware-based prefetching in a more tightly connected environments. Recent work \cite{johnnycache} has shown that it is possible to avoid cache conflicts in fast memory when fast memory is completely used as a cache for FAM. Our work employs only part of the fast memory as a cache. 
%The cost of page fault handling overheads has prompted advocating the movement of some of the O/S work into hardware \cite{mfoe,revisiting-page-walks}.

Data placement and movement can have a significant impact in a tiered memory systems and recent work \cite{tpp,nimble-page-management,google-lstm-page-lifetime,utility-based-hybrid-memory,pond} has considered strategies for keeping hot or more frequently utilized data in higher performance tiers. These strategies are complementary to our approach of mitigating the latency when the slower memory has to be accessed. Few of the tiering approaches require intimate knowledge of workload, to create hot/cold profile of application's memory footprint, which might not be possible for every kind of system use case.

PreFAM\cite{prefam} proposes prefetching at a distance from FAM into local DRAM cache. While there might be similarities in the system architecture, PreFAM doesn't consider resource contention at FAM, due to other participating nodes.

Direct CXL \cite{direct_cxl} implements a memory pooling solution leveraging CXL.mem protocol, that has access latency around 200 ns. DRAM cache prefetching along with bandwidth optimizations proposed in this work are agnostic to CXL fabric latency.

Recent work \cite{hotbox} has suggested that the employment of large pages may not be universally beneficial in disaggregated memory systems. Our work here considers the movement of sub-pages of data between DRAM and FAM to reduce the access latencies at FAM. 

\section{Conclusion}
This paper proposed a prefetching mechanism for caching sub-page blocks from FAM in a portion of the DRAM. The prefetcher learns from  LLC misses heading to FAM and issues prefetch requests to bring FAM data to DRAM to reduce the latency of future accesses. We considered two optimizations to mitigate congestion: compute-node based issue rate management mechanism based on observed latencies, and a memory-node based weighted fair queuing mechanism. We show that both of these mechanisms can improve the IPC of non-adaptive DRAM cache prefetch by 7-10\%. Our evaluation reveals that both of the approaches are effective in improving performance in different system configurations and workloads. 
\bibliographystyle{IEEEtranS}
\bibliography{refs}
\end{document}